# Scene-based field validation of wearable light loggers

Burcu Gemici[1, [0009-0003-7620-7390]]

Niloufar Tabandeh[2, 3, [0000-0002-7971-8589]]

Cansu Ozkucukler[1, [0009-0003-0876-6908]]

Johannes Zauner[3, [0000-0003-2171-4566]]

Altug Didikoglu[1,4, [ 0000-0002-5582-6956], *]

Manuel Spitschan[2, 3, 5, 6, [0000-0002-8572-9268], *, #]

[1] Izmir Institute of Technology, Department of Neuroscience, Izmir, Türkiye

[2] Max Planck Institute for Biological Cybernetics, Translational Sensory & Circadian Neuroscience, Tübingen, Germany

[3] TUM School of Medicine and Health, Technical University of Munich, Munich, Germany

[4] Centre for Biological Timing, Division of Diabetes Endocrinology and Gastroenterology, School of Medical Sciences, Faculty of Biology Medicine and Health, University of Manchester, Manchester, UK

[5] TUM Institute for Advanced Study (TUM-IAS), Technical University of Munich, Garching, Germany

[6] TUMCREATE Ltd., Singapore, Singapore

* Equal contributions

# Correspondence: Manuel Spitschan, Technical University of Munich, Am Olympiacampus 11, 80809 Munich, Germany, manuel.spitschan@tum.de

## Abstract

Wearable light loggers are increasingly used to measure personal light exposure. However, there is no standardised method to test how well these devices perform in real-world settings. To address this, we developed a scene-based field validation framework combining wearable light loggers, laboratory-grade spectral reference instruments, and image-based scene characterisation to evaluate wearable performance in the field. We applied the validation framework to ActTrust2 and ActLumus light loggers using 433 natural, everyday scenes across two sites: Tübingen, Germany (n=210), and Izmir, Türkiye (n=223), spanning indoor and outdoor environments in daylight, artificial light and mixed scenarios, across wide range of photopic and melanopic equivalent daylight illuminances. Both light loggers exhibited high agreement with the reference instruments ($R^2$=0.988–0.990), but they systematically underestimated light exposure. The lighting condition, scene complexity, time of day, and study site contributed significantly to measurement bias. Resampling under specific lighting conditions indicated that bias ranged from −0.065 log units in daylight-only situations to −0.195 log units in artificial-only situations. This suggested that single-condition assessments with limited spectra or scene categories can underestimate or overestimate overall wearable performance. Bootstrap resampling demonstrated that performance estimates stabilised at approximately 100 scenes, indicating that a diverse sample of this size is sufficient for reliable field validation. Finally, cross-site validation showed consistent performance across two sites, supporting the framework's reproducibility. Overall, these findings establish our validation framework as an effective tool for guiding scene diversity and sampling design for ecologically valid field validation of wearable light loggers.

**Keywords:** wearable light loggers; light exposure; circadian rhythms; field validation; real-world settings; measurement accuracy

## Statement of significance

Wearable light loggers are becoming common tools in circadian, sleep and environmental health research for measuring real-world light exposure. Most validation procedures rely on laboratory-based characterisation techniques that are not reflective of real-world scenes that the loggers might measure in practice. This study introduces a novel scene-based field validation framework combining wearable light loggers, laboratory-grade spectral reference instruments, and image-derived scene characterisation across a variety of indoor and outdoor environments. The results show that lighting condition, time of day, study site, and the spatial complexity of the lighting environment all contribute to measurement bias, and that assessments based on a limited set of lighting conditions can substantially underestimate or overestimate overall device performance. Bootstrap resampling further demonstrated that a diverse selection of approximately 100 scenes is sufficient for stable and reliable performance estimates. Validated across two geographically distinct sites using locally available instrumentation, our framework provides an ecologically valid and reproducible framework to guide both scene selection and sample-size planning in future wearable light exposure research.

## Introduction

Wearable light loggers measure individual light exposure under real-world conditions while capturing the variability in environmental lighting, behaviour, and sociocultural factors [1–3]. These devices have significantly advanced our understanding of how short-term light exposure and long-term circadian light–dark cycles influence human health and well-being, including sleep [4, 5], alertness [5], mood [6], and metabolic health [7], and have become essential tools for studying exposure to light and optical radiation in the context of health in general. The reliability and comparability of these findings, however, depend on the characteristics and performance of the measurement devices used. Currently, there are no standard or harmonised off-the-shelf solutions to assess the quality of wearable light measurements in the field.

Wearable light loggers encompass a wide range of technologies, from expensive research-grade commercial devices to inexpensive open-source systems designed for implementation in variety of environments [8]. This diversity is accompanied by substantial heterogeneity in device characteristics, including sensor design, spectral resolution, angular response, dynamic range, and overall measurement performance, as well as differences in calibration practices such as the choice of light sources, reference standards, and validation procedures [9]. In a recent survey of more than 50 wearable contemporary light loggers, van Duijnhoven et al. reported considerable variability in device characteristics and usability, along with limited and inconsistent reporting of performance and calibration data [9]. The authors highlighted the need for improved standardisation, reporting, and validation practices to enable more reliable use of these devices in research.

Several prior studies have also evaluated the performance and calibration of wearable light loggers in both laboratory and field settings, focusing on the comparison of a small number of devices at a time. In a study by Figueiro et al., differences in light measurements have been reported between the Daysimeter and Actiwatch Spectrum devices in participant-based protocol, and laboratory comparisons with a calibrated illuminance meter showed that the Actiwatch

Spectrum exhibited higher cosine response and spectral mismatch errors and greater variability in recorded illuminance values across units than the Daysimeter [10]. A later study by Markvart et al., investigating different calibration methods demonstrated that Actiwatch Spectrum devices showed substantial inter-unit error and tended to overestimate photopic illuminance relative to a calibrated photometer. The authors proposed a side-by-side field calibration using a cosine-corrected photometer under diffuse overcast sky conditions, where multidirectional sunlight reduces inter-device variability and measurement error [11]. More recently, Ishihara et al. evaluated the performance of LYS, ActLumus, GENEActiv, and ActiGraph wGT3X-BT devices against a criterion spectrometer under controlled laboratory conditions and found substantial variation in photopic illuminance and melanopic equivalent daylight illuminance (mEDI) measurements across device types. A subsequent free-living comparison between the ActLumus and ActiGraph devices revealed marked differences in recorded illuminance, with the greatest discrepancies occurring at light levels ≤100 lux, consistent with the laboratory findings [12]. These findings suggest that manufacturer specifications alone are insufficient to guarantee reliable measurements under real-world conditions and highlight the need for further technical standardisation. Collectively, these studies emphasise the importance of rigorous evaluation and, where appropriate, field validation of wearable light loggers in research, particularly given the inter-device variability and the dynamic spectral and spatial characteristics of real-world lighting environments.

Importantly, previous studies have often evaluated light logger performance in constrained laboratory settings that are not representative of real-world deployment conditions, thereby creating a mismatch between characterization and actual measurement conditions. This alignment is important because device characterization is only informative to the extent that it predicts performance under actual deployment conditions; otherwise, laboratory estimates of accuracy may not generalize to real-world use, potentially biasing exposure assessment.

Here, we developed and validated a novel framework to characterize the performance of wearable light loggers under real-world settings with respect to scenes that may be encountered by an active user. Our framework comprises a standardised field-validation protocol and analysis pipeline integrating wearable light loggers, laboratory-grade spectral reference instruments, and image-based scene characterisation under ecologically valid conditions. Specifically, our objectives were to (1) quantify the agreement of two wearable light loggers with laboratory-grade spectral reference instruments across a diverse set of naturalistic scenes; (2) determine whether measurement bias varies as a function of lighting and scene characteristics (3) evaluate the robustness of performance estimates across varying sample sizes and lighting condition compositions; and (4) assess whether the framework can be successfully reproduced across geographically and environmentally distinct sites using locally available instrumentation.

## Methods

### Overall design

We collected measurements using a structured framework, named SceneVAL (see **Supplementary Protocol**), to assess the performance of the wearable light loggers under real-world conditions. Two research-grade wearable light loggers were deployed in parallel with laboratory-grade spectral reference instruments and action cameras at two sites, Tübingen, Germany and Izmir, Türkiye; collecting light exposure measurements across a wide variety of lighting and scene characteristics to assess device agreement and sources of measurement bias. All instruments were mounted on a fixed tripod rig, ensuring that wearable light loggers and reference instruments sampled the same lighting environment in each scene. Action cameras were used to obtain spatial and structural information for each scene. The two sites differed in latitude, season, and built environment, with Germany representing an urban Northern European context and Türkiye representing a Mediterranean university campus and suburban setting. Site-specific reference instruments were used, reflecting differences in available laboratory equipment across locations. This multi-site design allowed the evaluation of the consistency in independent deployments of the SceneVAL framework with different instrumentation and environmental conditions.

## Measurement setup

The instrumentation was mounted on a tripod rig equipped with a horizontal rail, accommodating a laboratory-grade spectral reference instrument, wearable light loggers, and a GoPro action camera. All devices were positioned at fixed offsets relative to the reference sensor (**Figure 1A**; see **Supplementary Protocol, Steps 1–2**). Reference instruments and action cameras were site-specific: a JETI Spectraval 1501 spectroradiometer (JETI Technische Instrumente GmbH, Jena, Germany) and GoPro HERO10 Black (GoPro, San Mateo, CA, USA) were used in Tübingen, and a Konica Minolta CL-500A spectrophotometer (Konica Minolta, Tokyo, Japan) and GoPro HERO8 Black (GoPro, San Mateo, CA, USA) in Izmir. The ActTrust 2 (Condor Instruments, São Paulo, Brazil) was the same device across both sites, while a different ActLumus unit (Condor Instruments, São Paulo, Brazil) was used at each site. Both wearable devices recorded photopic illuminance; ActLumus additionally recorded the melanopic equivalent daylight illuminance (mEDI), a biologically relevant metric for circadian-effective light exposure[12.] The GoPro HERO10 Black and GoPro HERO8 Black were operated in Wide mode, with nominal horizontal and vertical fields of view of 121°/93° and 122.6°/94.4°, respectively.

To assess potential shadowing between devices (see **Supplementary Protocol, Step 3a**), we geometrically quantified the fraction of the upper hemisphere occluded by neighbouring instruments for each sensor configuration. Using the corner coordinates of the devices, azimuth and elevation angles were calculated relative to the observer position for each of them. A 360° azimuth skyline was then constructed, in which the maximum elevation angle subtended by any device was retained for each azimuth direction. The total obstructed solid angle was obtained by integrating the skyline over the hemisphere and expressing the result as a fraction of the upper hemisphere (2π sr), thereby accounting for overlapping obstructions between devices.

In addition, the cosine-weighted fraction of the upper hemisphere subtended by the action camera field of view (FOV) was estimated to characterise the proportion of the light environment captured by the scene recordings (see **Supplementary Protocol, Step 3b**). The horizontal and vertical FOV of the camera were converted into image-plane half-angle bounds, and the cosine-weighted solid-angle element was integrated numerically over these bounds, and expressed relative to the total cosine-weighted solid angle of the upper hemisphere (π sr).

**Sampling and measurement procedure**

Data collection followed the SceneVAL field protocol (see **Supplementary Protocol, Step 4**). It was conducted in Tübingen, Germany (48.5° N, 9.1° E) during August and September 2025 and in Izmir, Türkiye (38.3° N, 26.6° E) during November 2025. In this study, each "scene" was a single measurement event at a fixed tripod position and orientation under specific lighting conditions. These scenes were recorded between 08:00 and 23:00 local time and included various indoor and outdoor scenarios containing illumination by both daylight and artificial light, as well as mixtures. Where feasible, the rig was rotated 180° to capture the lighting environment in the opposite direction. Throughout each measurement day, the ActTrust 2 and ActLumus devices recorded continuously at 1-second resolution. For each defined scene, the reference instrument and GoPro camera were manually triggered in succession, with all device clocks synchronised manually prior to each session.

**Data management**

Data processing and management were performed using custom made scripts developed in Python (v3.12), integrating reference instrument and wearable light logger data, and GoPro images via synchronised timestamps within a structured directory system. Each scene was annotated with metadata describing measurement time, geographic coordinates, environmental conditions, scene characteristics, and protocol information (see **Supplementary Protocol, Step 5**). Wearable light logger data were temporally aligned with reference instrument measurements by averaging values across the corresponding acquisition intervals. When integration-time metadata from the reference instrument was unavailable, the exposure window was approximated as a 3-second interval relative to the recorded reference timestamp.

**Computation of photometric and α-opic metrics**

For the measurements in Izmir, spectral data from the CL-500A spectrophotometer were converted into photopic and α-opic quantities in R (v4.4.3) [14], using the alphaopics package [15] adhering to the International Standard CIE S 026/E:2018 [16]. Photopic illuminance and mEDI were derived by integrating spectral data with the spectral weighting functions.

**Image-derived scene characteristics**

Image-based scene metrics were computed to characterise the spatial structure of each lighting scene and examine whether scene complexity independently predicts wearable measurement bias. Wearable sensors integrate light over the upper hemisphere through a fixed cosine-corrected optic, whereas reference instruments measure from a defined point and direction; in spatially heterogeneous scenes, the angular distribution of light reaching each sensor may differ systematically, producing discrepancies that are not purely a function of illuminance magnitude and that these spatial metrics are designed to capture. Raw GoPro images were converted to DNG format using Adobe DNG Converter and processed in Python using scikit-image [17] as grayscale images. Three metrics were derived: RMS contrast (standard deviation of grayscale pixel intensities), Shannon entropy (grayscale intensity distribution complexity), and edge density (mean Sobel gradient magnitude).

**Scene classification**

To disaggregate by scene type, we collected additional metadata on each scene. Each scene was labeled with contextual descriptors, including the scene type (e.g., i*ndoor kitchen*, *outdoor grove*, full list provided in the **Supplementary Protocol, Step 5**), lighting condition (*indoor* or o*utdoor*; *daylight*, *artificial light*, or *mixed*), and study site (Germany or Türkiye). Solar parameters, including sunrise, sunset, and photoperiod, were calculated in R, using geographic coordinates and timestamps with the lubridate [18] and suncalc [19] packages. A categorical time of day label was then assigned to each scene based on its timing relative to local sunrise and sunset. Scenes occurring before sunrise were classified as *pre-dawn*, those occurring after sunset as *after sunset*, and those occurring between sunrise and sunset were classified according to the fraction of the photoperiod elapsed since sunrise: *morning* (≤50%), *midday* (>50–75%), and *evening* (>75–100%).

**Statistical analysis**

All statistical analyses were performed using R, with the following packages: stats [20], car [21], broom [22], corrplot [23], emmeans [24], purrr [25], ggplot2 [26], and ggdist [27]. Descriptive statistics were reported as ranges for continuous variables and as frequencies (n, %) for categorical variables. Two-tailed p-values were used for hypothesis testing with a significance level set to 0.05. We did not apply corrections for multiple testing because the objective was to detect potential systematic differences between the light logger and the reference measurement, rather than to establish significant effects across multiple independent endpoints. Since a null difference between the devices is the desired result in this validation context, multiplicity correction would inappropriately reduce statistical power and increase the risk of false-negative findings.

Scatterplots with regression lines and Bland-Altman plots were used to visualize the differences in log-transformed mEDI and photopic illuminance between devices. The measurement error for both metrics was defined as the difference between wearable and reference log-transformed readings (hereafter referred to as bias). The predictors of mEDI bias for the ActLumus device were analyzed using a linear model, including study site, lighting condition, time of day, and spatial image metrics derived from GoPro images; correlation matrices and variance inflation factors (VIFs) were observed to check collinearity. Type III ANOVA was applied to the fitted model, using effect sizes expressed as partial eta squared and standardized regression coefficients to weigh the contributions of numeric predictors. Post hoc analyses were conducted for categorical predictors: estimated marginal means (EMMs) were computed for study site, lighting condition, and time of day, with pairwise comparisons adjusted for multiple testing via Tukey's method. Bias distributions across lighting conditions and study sites were visualized using raincloud plots, combining density estimates, boxplots, and individual data points.

The performance of the wearable light loggers and the stability of the performance assessment were evaluated through bootstrap resampling [28]. Bootstrapping was performed on the pooled dataset without stratification by scene type, lighting condition, site or time of day. Linear models were fitted to log-transformed reference and wearable mEDI values within each resampled dataset, from which $R^2$ and root-mean-square error (RMSE) were derived, alongside bias. These metrics were evaluated across varying sample sizes (10 to 1000) and four fixed-size (n=50) lighting scenarios – *all-indoor, all-outdoor, daylight-only, and artificial-only* with resampling within each scenario performed without further stratification. Finally, cross-site validation examined whether the SceneVAL framework produces replicable performance estimates when independently deployed at a geographically distinct site with different instrumentation, by training models on one site and testing on the other. Linear models trained on data from one site (Germany or Türkiye) predicted wearable mEDI for the other, with performance assessed through $R^2$, mean absolute error (MAE), and mean bias.

## Results

### Measurement rig for wearable light logger validation performs with negligible inter-device obstruction across sites

Our measurement rig was successfully deployed at both measurement sites using locally available instrumentation, demonstrating that the configuration is not dependent on identical equipment or fixed inter-device distances. Geometrical quantification of mutual shadowing confirmed that neighbouring instruments occluded only a small fraction of the upper hemisphere at each site. At the Türkiye site, occlusion values were 3.37% for the reference instrument and 0.92% for the ActTrust 2; at the Germany site, corresponding values were 0.20% and 0.53%, respectively. Across both configurations, inter-device shadowing was negligible, confirming that the rig design did not meaningfully compromise measurement integrity.

**A new diverse real-world scene database for wearable calibration**

A total of 433 scenes were captured across Tübingen, Germany (n=210) and Izmir, Türkiye (n=223), spanning indoor and outdoor environments from 08:00 to 23:00 with similar temporal and lighting condition coverage across both sites (**Figure 1F–G**). Light levels covered a wide dynamic range, with photopic illuminance extending from 0.000213 to 89,763 lux in Germany and 0.359 to 93,745 lux in Türkiye, and mEDI reaching maximum values of 77,485 lux and 81,539 lux, respectively. Photoperiods ranged from 11.8 to 13.8 hours in Germany and 9.8 to 10.5 hours in Türkiye (**Table 1**). Together, our dataset provides sufficient diversity in lighting conditions, scene types, and geographic contexts to support agreement analysis, bias modelling, and reproducibility testing.

**Wearable light loggers closely track reference measurements but systematically underestimate light exposure**

Both wearable light loggers showed strong agreement with the reference instrument across the full dynamic range of observed light levels, though deviations were more pronounced at lower intensities (**Figure 2**). The ActLumus demonstrated closer agreement with the reference for both mEDI (**Figure 2A**) and photopic illuminance (**Figure 2B**) than the ActTrust 2 (**Figure 2C**). Despite this overall strong agreement, both devices exhibited a consistent negative bias, systematically underestimating light levels relative to the reference.

For the ActLumus device, mEDI bias was not uniform across conditions. Type III ANOVA revealed significant main effects of site ($p=0.006$, partial $\eta^2=0.018$), lighting condition ($p<0.001$, partial $\eta^2=0.061$), and time of day ($p=0.025$, partial $\eta^2=0.022$). Across lighting conditions, *outdoor artificial light* produced the strongest underestimation (EMM=−0.26, SE=0.04), corresponding to approximately a 45% reduction in linear scale relative to the reference, and this was significantly greater than under *indoor artificial light* (EMM=−0.11, SE=0.02; $p=0.005$), *indoor mixed* (EMM=−0.08, SE= 0.02; $p=0.001$), and *outdoor daylight* (EMM=−0.08, SE=0.02; $p=0.002$). Across time of day categories, underestimation was strongest after *sunset* (EMM=−0.17, SE=0.03), corresponding to approximately a 33% reduction relative to the reference, and was significantly greater than in the *morning* (EMM=−0.07, SE=0.02; $p=0.037$), whereas all other time of day comparisons were non-significant. Notably, image-derived scene metrics, RMS contrast, edge density, and Shannon entropy, were independently associated with bias magnitude ($\beta=-0.033$, −0.035, and 0.026, respectively; **Table 2**). These patterns are illustrated in **Figure 3**, where *outdoor artificial light* stands out with the largest bias and greatest variability, in contrast to the closer agreement observed under daylight conditions.

**Performance estimates are robust across sampling strategies and generalize across sites**

Bootstrap subsampling across simulated sample sizes from n=10 to 1000 indicated that mean bias (−0.105 to −0.107 log units) and mean $R^2$ (0.988–0.990) remained stable regardless of sample size (**Table S3**). RMSE stabilised at approximately 0.14 log units from $n\geq100$, suggesting that around 100 scenes are sufficient to obtain reliable performance estimates, with larger samples improving precision but not meaningfully changing overall accuracy.

The lighting composition of the validation sample also mattered. At a fixed sample size of 50, daylight-only scenes yielded the smallest bias (−0.065 log units) and closest agreement with the reference, whereas electric-only scenes showed the strongest underestimation (−0.195 log units) and lowest $R^2$ (0.967; **Table S4**).

Cross-site validation demonstrated that the SceneVAL framework produces replicable performance estimates across geographically and environmentally distinct deployments. Models trained on one site and tested on the other maintained high predictive performance ($R^2$=0.988–0.989) with low mean absolute error (0.099–0.121 log units; **Table S5**). Mean bias values were small but opposite in direction between sites (−0.069 and +0.070 log units).

**Table 1. Descriptive statistics of the measurements (n=433).**

| Site | Tübingen/Germany (n=210) | Izmir/Türkiye (n=223) |
|---|---|---|
| **Geographical coordinates (°)** | 48.5 N, 9.1 E | 38.3 N, 26.6 E |
| **Measurement period** | 26 Aug – 29 Sep 2025 | 4 Nov – 26 Nov 2025 |
| **Photoperiod (h), range** | 11.8 – 13.8 | 9.8 – 10.5 |
| **Reference instrument model** | JETI spectraval 1501 spectroradiometer | KONICA MINOLTA CL-500A spectrophotometer |
| **Photopic illuminance (lux), range** | 2.13e-04 – 89,763 | 0.359 – 93,745 |
| **mEDI (lux), range** | 3.26e-04 – 77,485 | 0.0892 – 81,539 |
| **Scene lighting, n (%)** | | |
| *Indoor daylight* | 53 (49.1%) | 55 (50.9%) |
| *Indoor artificial light* | 38 (46.3%) | 44 (53.7%) |
| *Indoor mixed* | 28 (46.7%) | 32 (53.3%) |
| *Outdoor daylight* | 71 (49.3%) | 73 (50.7%) |
| *Outdoor artificial light* | 18 (50.0%) | 18 (50.0%) |
| *Outdoor mixed* | 2 (66.7%) | 1 (33.3%) |

*Abbreviations: mEDI: melanopic equivalent daylight illuminance.*

**Table 2. Type III ANOVA results for the linear bias model.**

| Predictor | Df | Sum Sq | F value | p value | Partial η² | Std. Beta |
|---|---|---|---|---|---|---|
| Site | 1 | 0.182 | 7.602 | **6.08e-03** | 0.018 | - |
| Lighting | 5 | 0.655 | 5.479 | **6.72e-05** | 0.061 | - |
| Time of day | 3 | 0.226 | 3.145 | **0.025** | 0.022 | - |
| RMS contrast | 1 | 0.320 | 13.401 | **2.84e-04** | 0.031 | 0.033 |
| Entropy | 1 | 0.155 | 6.489 | **0.011** | 0.015 | -0.026 |
| Edge Density | 1 | 0.255 | 10.659 | **1.19e-03** | 0.025 | 0.035 |

*Abbreviations: RMS: root mean square, $\eta^2$: partial eta squared, Std. Beta=standardized regression coefficient.*

**Table S3. Validation robustness across simulated sample sizes.**

| Sample size | Mean bias | SD bias | Mean R² | SD R² | Mean RMSE |
|---|---|---|---|---|---|
| 10 | -0.107 | 0.059 | 0.990 | 0.012 | 0.108 |
| 30 | -0.105 | 0.033 | 0.989 | 6.71e-03 | 0.131 |
| 50 | -0.106 | 0.026 | 0.988 | 5.16e-03 | 0.138 |
| 100 | -0.105 | 0.018 | 0.988 | 3.44e-03 | 0.140 |
| 250 | -0.106 | 0.012 | 0.988 | 2.26e-03 | 0.144 |
| 500 | -0.105 | 8.13e-03 | 0.988 | 1.56e-03 | 0.144 |
| 1000 | -0.105 | 5.65e-03 | 0.988 | 1.11e-03 | 0.145 |

*Abbreviations: RMSE: root mean square error. Bias refers to the difference between wearable and reference log-transformed values.*

**Table S4. Performance metrics across different validation strategies (fixed sample size N=50).**

| Strategy | Mean bias | SD bias | Mean R² | SD R² | Mean RMSE |
|---|---|---|---|---|---|
| *All-indoor* | -0.120 | 0.023 | 0.978 | 7.39e-03 | 0.116 |
| *All-outdoor* | -0.086 | 0.029 | 0.989 | 5.03e-03 | 0.146 |
| *Daylight-only* | -0.065 | 0.021 | 0.989 | 5.30e-03 | 0.123 |
| *Artificial-only* | -0.195 | 0.030 | 0.967 | 0.015 | 0.148 |

*Abbreviations: RMSE: root mean square error.*

**Table S5. Cross-site model validation results.**

| Training site | Testing site | Cross-R² | MAE | Bias |
|---|---|---|---|---|
| Germany | Türkiye | 0.989 | 0.099 | -0.069 |
| Türkiye | Germany | 0.988 | 0.121 | 0.070 |

*Abbreviations: MAE: mean absolute error.*

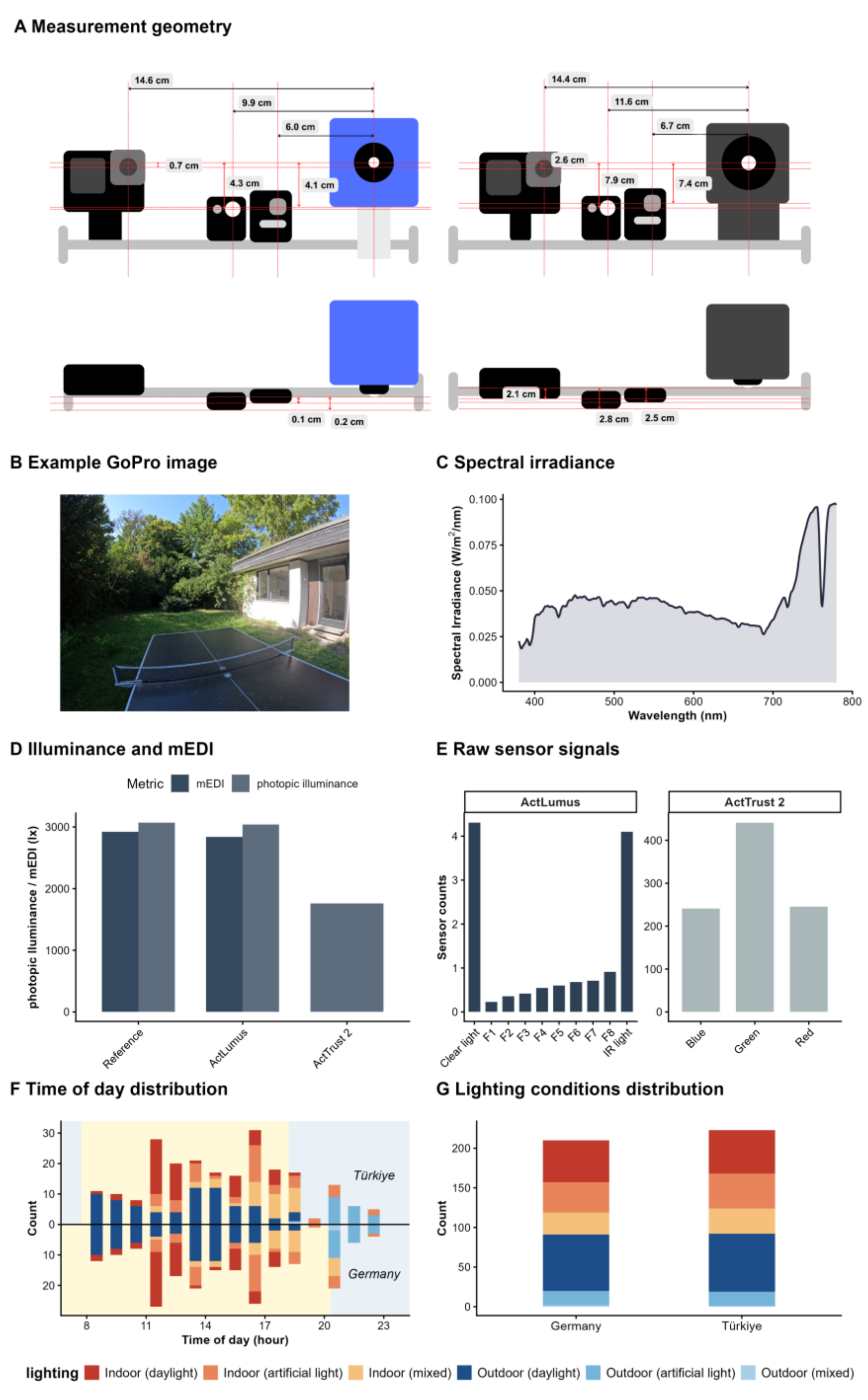

**Figure 1. Experimental setup and dataset characteristics.** (A) Schematic representation of the tripod-mounted measurement rigs used in Tübingen, Germany (left) and İzmir, Türkiye (right), showing the spatial alignment and distances (cm) between the reference instrument, GoPro camera, ActTrust 2, and ActLumus sensors. (B) Example GoPro image of an outdoor scene in Tübingen (Recorded: 2025-08-26 10:34). For the same example scene, the corresponding: (C) spectral irradiance (W/m²/nm) recorded by the reference instrument; (D) photopic illuminance (lx) and mEDI (lx) values calculated by the evaluated devices; and (E) raw sensor signals from the ActLumus (10-channel multispectral) and ActTrust 2 (RGB) devices. (F) Distribution of total scenes by time of day and lighting condition for both sites. Background shading indicates the photoperiod variation across all sampling dates for each site. Yellow-shaded regions represent the range of core daylight hours (from earliest recorded sunrise to latest recorded sunset), while the blue-shaded regions indicate nighttime. (G) Distribution of total scenes by lighting condition for both sites.

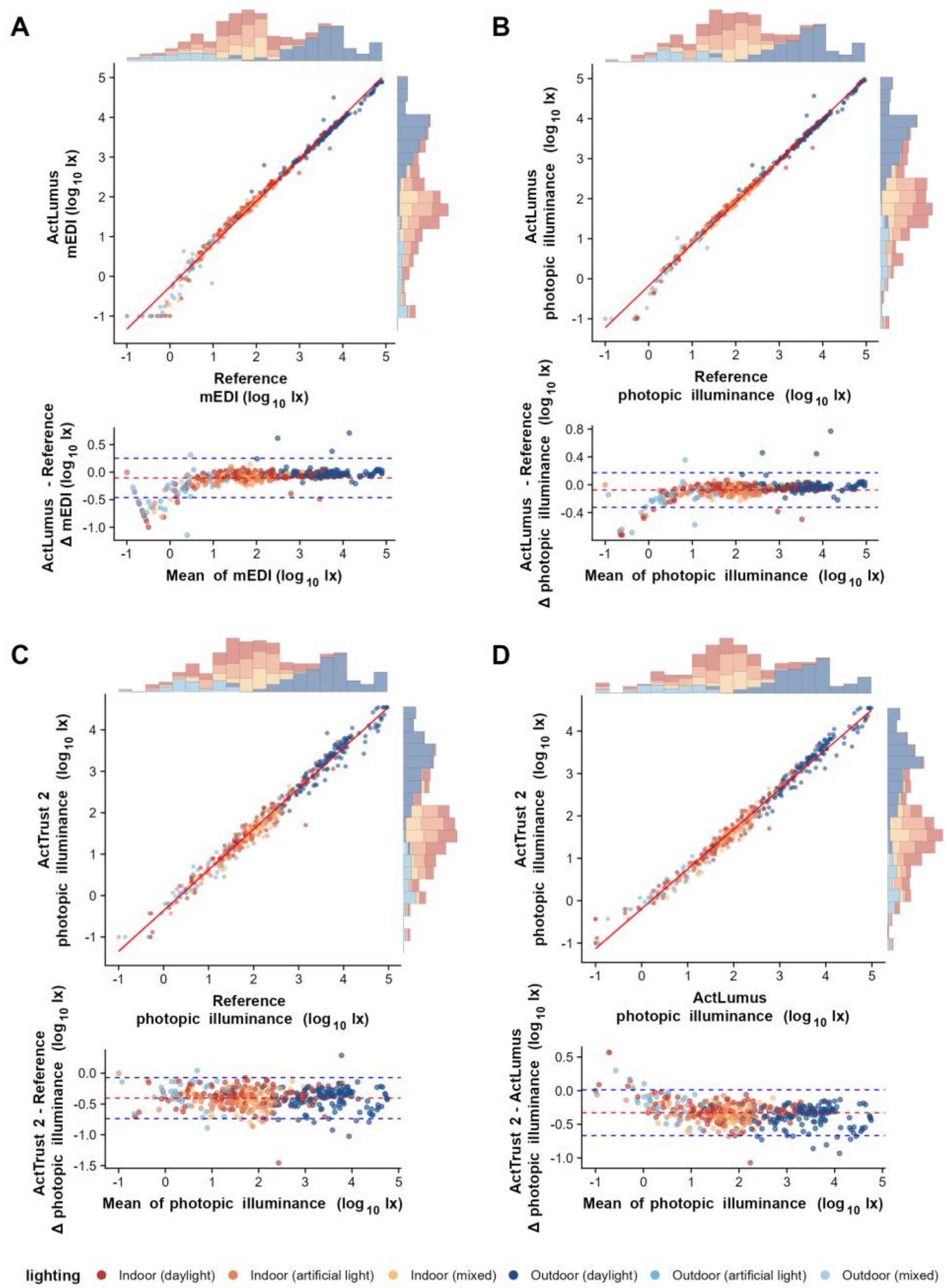

**Figure 2. Performance comparison of wearable light loggers against reference instrument.** Scatterplots (top panels) and Bland–Altman plots (bottom panels) showing agreement between devices for photopic illuminance and mEDI (log10-transformed, lx). (A) Reference instruments. ActLumus for mEDI. (B) Reference instrument vs. ActLumus for photopic illuminance. (C) Reference vs. ActTrust 2 for photopic illuminance. (D) ActLumus vs. ActTrust 2 for photopic illuminance. Scatterplots include the linear regression fit (solid red line with shaded 95% confidence interval). In the Bland–Altman plots, the y-axis represents the difference (Δ) between the two devices, showing the mean bias (dashed

red line) and 95% limits of agreement (dashed blue lines). Points are colour-coded by lighting condition (indoor/outdoor and daylight/artificial light/mixed). Marginal histograms represent the pooled distribution of all samples.

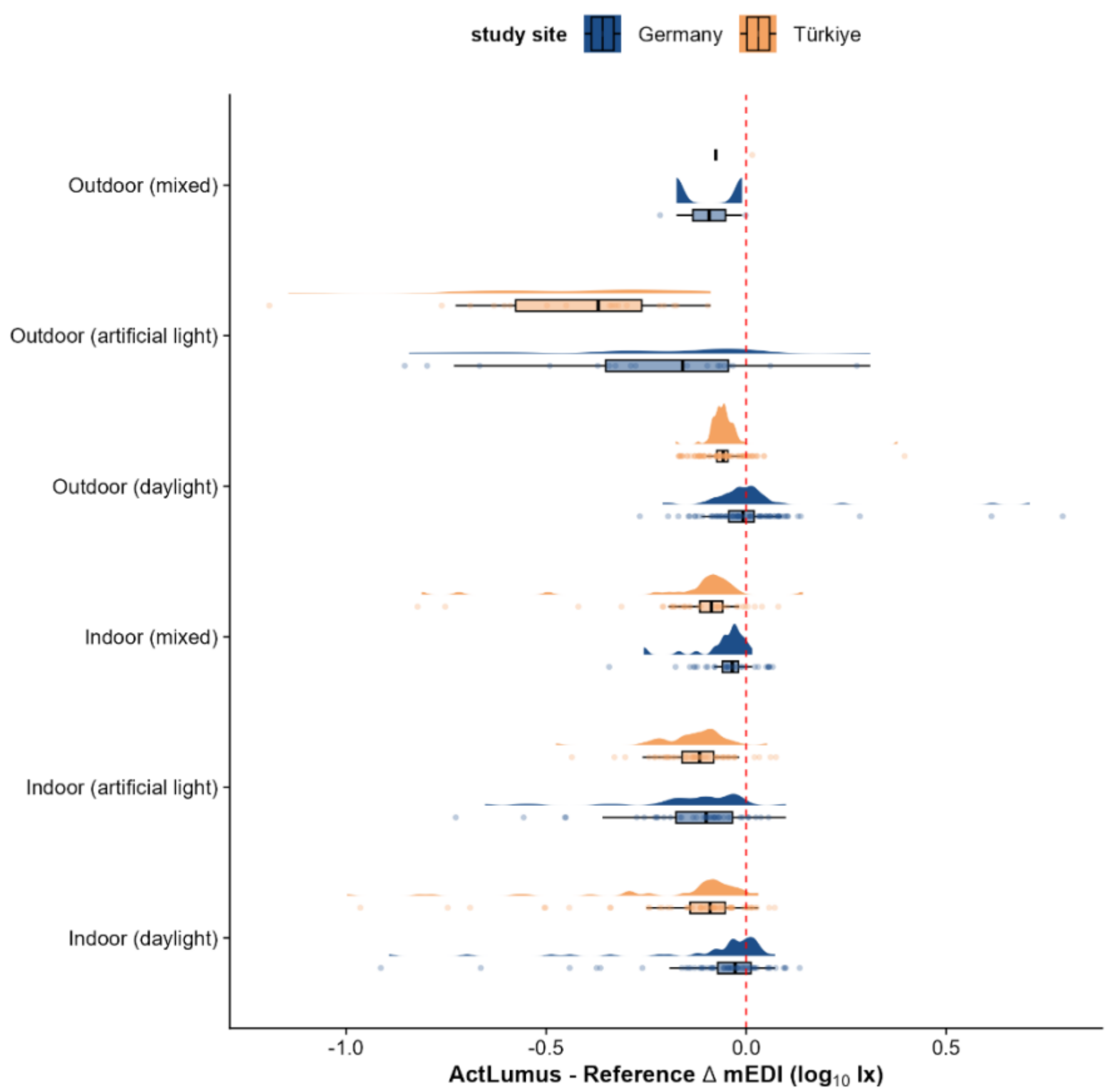

**Figure 3. Distribution of the difference (Δ) in mEDI (log10-transformed, lx) between the ActLumus and reference instrument by site and lighting condition.** Data are grouped by lighting condition and partitioned by study site: Germany (blue) and Türkiye (orange). The central dashed red line at 0.0 represents perfect agreement.

## Discussion

Wearable light loggers are increasingly deployed to quantify personal light exposure in studies linking light to circadian, metabolic, and mental health outcomes. Ensuring that these devices perform accurately across the diverse conditions of real-world use is therefore a prerequisite for drawing valid inferences from field data. Yet, prior to this work, there was no standardised framework for evaluating their performance under naturalistic conditions. This study introduced and evaluated a novel framework for validating wearable light loggers based on various natural scenes in real-world, everyday settings. It was applied across 433 scenes at two different sites: Tübingen, Germany (n=210), and Izmir, Türkiye (n=223). Both tested devices, the ActLumus and ActTrust 2, showed strong agreement with laboratory-grade reference instruments across a wide dynamic range of photopic illuminance and mEDI. The ActLumus demonstrated closer agreement than the ActTrust 2, consistent with prior findings that ActLumus shows higher accuracy than other wearable light sensors [11]. However, both devices displayed a consistent negative bias, systematically underestimating light levels relative to the reference. This bias was not uniform for the mEDI readings of the ActLumus device: it varied significantly as a function of lighting condition, time of day, and image-derived scene characteristics. The strongest underestimation was observed under *outdoor artificial light* conditions, while *daylight* conditions showed the closest agreement. This pattern is likely primarily driven by the lower absolute mEDI levels of *outdoor artificial light* conditions, given that both devices showed greater deviations at lower light intensities. Additionally, the spatial heterogeneity of directional artificial light sources may further contribute to this discrepancy by creating a greater angular mismatch between the wearable sensor's cosine-weighted integration and the incident light field.

A similar ambiguity applies to the time of day effect: lower underestimation in the *morning* than *after sunset* may reflect differences in the angular distribution and spatial structure of lighting environments across the day, but *morning* conditions also tend to involve systematically higher light levels than *after sunset* environments. In both cases, light level, spatial structure, and temporal context are confounded in the present dataset and cannot be fully disentangled without experimental designs that independently vary between these factors. Regardless of the underlying mechanism, the practical implication is clear: wearable measurement error is systematically elevated under the artificial lighting environments that typify evening and nighttime life, precisely the conditions where accurate characterisation may matter most, given that the human circadian system is highly sensitive to relatively low levels of evening light exposure and even modest illuminance can substantially affect circadian timing and melatonin regulation [29]. Systematic underestimation in these conditions could bias studies seeking to characterize light exposure and its effects on circadian timing and health. Such condition-dependent errors are difficult to identify through laboratory validation alone and illustrate the value of our framework for assessing device performance in real-world settings.

The image-derived scene characteristics were independently associated with bias magnitude, though they were not uniform in direction. Higher RMS contrast and edge density were associated with greater underestimation ($\beta$=0.033 and 0.035, respectively), while higher Shannon entropy was associated with smaller underestimation ($\beta$=−0.026). This divergence in the direction of associations across those metrics suggests that they capture distinct aspects of scene structure, which may relate differently to measurement error. Consequently, a single unified explanation based solely on spatial complexity is difficult to establish. Moreover, *outdoor artificial light* scenes, which showed the strongest underestimation, did not exhibit the highest values on these metrics; *outdoor daylight* scenes showed higher mean RMS contrast, entropy, and edge density. Together, these patterns suggest that the image metrics reflect a source of variance in bias that is independent of the lighting condition effect. Notably, although RMS contrast, entropy, and edge density have frequently been used to characterise visual scene complexity and clutter in perception research, these metrics were not developed to model the optical characteristics that determine light sensor performance. Whether and how such metrics translate to predicting measurement error in wearable light loggers, whose response is governed by directional sensitivity, aperture geometry, and spectral weighting rather than perceptual salience, remains an open question. Developing image-based descriptors specifically suited to characterizing the optical environment as experienced by a sensor, rather than by a human observer, represents an important methodological gap that future work should address.

Prior studies evaluating wearable light logger performance have compared devices under a range of settings but have not validated wearable measurements against a laboratory-grade reference instrument under naturalistic, real-world conditions. Across all three studies, device performance under naturalistic conditions was either inferred from laboratory findings or assessed relative to another wearable, leaving open the question of how accurately these devices measure light when deployed in the diverse and spatially complex environments of everyday life. The present study addresses this gap directly. By deploying laboratory-grade spectral reference instruments alongside wearable light loggers across 433 naturalistic scenes, our framework provides ground-truth validation of wearable performance under the actual conditions of real-world use. Notably, while Markvart et al. recommended calibrating under diffuse daylight conditions to minimise inter-device variability, our results show that *daylight-only* validation produces the smallest bias estimate (−0.065 log units) compared to *artificial-only* conditions (−0.195 log units), meaning that a *daylight-only* validation protocol, however well-controlled, will substantially overestimate overall device performance across the full range of real-world lighting conditions.

The scene-based design of our framework offers several concrete advantages. The dataset was deliberately sampled to ensure diverse indoor and outdoor environments, spanning over five log units of dynamic range and balanced lighting conditions across both sites, effectively representing real-world wearable deployment scenarios. The inclusion of GoPro image capture alongside light measurements enabled post-hoc investigation of the structural factors driving bias, a dimension entirely inaccessible in purely photometric validation approaches. Bootstrap resampling further showed that approximately 100 diverse scenes are sufficient for stable and reliable performance estimates, providing actionable guidance for future implementations and suggesting that scene diversity, spanning a wide dynamic range of light levels and a representative mix of lighting conditions, matters more than sheer sample size.

Cross-site validation provided an additional test of the framework's robustness and reproducibility. The site effect observed in the bias model and the opposite-direction cross-site bias values are best understood together. Within-site, the wearable underestimated more strongly against the Konica Minolta in Türkiye than against the JETI Spectraval in Germany; cross-site, models trained on Germany data underestimated Türkiye values while models trained on Türkiye data overestimated Germany values by almost exactly the same magnitude (~0.07 log units). The magnitude of this cross-site difference is comparable to the bias observed under *daylight-only* conditions (−0.065 log units), placing it within the range of condition-dependent variability in this dataset. Taken together, both patterns are consistent with a systematic offset between the two reference instruments, most likely reflecting differences in their absolute calibration. While a contribution from genuine geographic or environmental factors cannot be ruled out, it remains indistinguishable from instrument effects due to the full confounding between site and reference instrument in the present design. Nevertheless, near-perfect cross-site $R^2$ values indicate a highly consistent functional relationship between wearable and reference measurements across deployments. This supports the reproducibility of the framework's performance estimates despite differences in geographic location, season, rig configuration, and reference instrument used.

Beyond validation, the scene-based approach may also enable the development of device-level correction factors based on scene-specific bias estimates. Because the framework characterises performance across a representative range of lighting conditions, the resulting error profiles could serve as the basis for post-hoc calibration, adjusting wearable measurements as a function of estimated lighting condition or intensity range. This would represent a meaningful advance over single-point laboratory calibration, which cannot account for the condition-dependent error structure documented here.

The *outdoor artificial light* condition also warrants particular attention in the context of site effects. As illustrated in **Figure 3**, this condition showed not only the largest absolute bias but also the greatest variability across scenes, a pattern that may partly reflect the comparatively smaller number of scenes collected under this condition relative to indoor and daylight categories. With fewer observations, condition-level estimates are inherently more sensitive to the specific scenes sampled, including the particular mix of lamp types present at each site. If the spectral composition of outdoor artificial sources differed systematically between Tübingen and Izmir, for instance, due to differences in street lighting infrastructure, this could contribute to the elevated cross-site variability observed in this condition, in a way that cannot be disentangled from the instrument confound described above.

Several limitations of the present study should be acknowledged. Most importantly, study site and reference instrument model were fully confounded in the present design (the JETI Spectraval 1501 was used exclusively in Germany and the Konica Minolta CL-500A exclusively in Türkiye) meaning that the observed between-site differences in bias are consistent with instrument-level calibration differences and cannot be separated from geographic or environmental effects. Researchers implementing our framework across multiple sites should therefore use a single reference instrument model or include a cross-instrument calibration step to enable direct comparison of absolute bias estimates across locations. Second, while the GoPro-derived image metrics provided useful proxies for scene complexity, they capture only a restricted fraction of the cosine-weighted hemispherical light field (~67–68%, based on field of view geometry) and are grounded in spatial statistics developed for human visual perception rather than sensor optics. As discussed above, whether such metrics adequately characterise the optical environment as experienced by a wearable light logger, particularly in terms of directional light distribution and point-source geometry, remains an open question. Future work could explore fisheye or full-hemisphere imaging alongside sensor-oriented descriptors that better capture the spatial distribution of light reaching the device. Third, the dataset was collected across two sites in Europe and Western Asia and may not fully represent the range of built environments, cultural lighting practices, and climatic conditions found globally; extending the framework to additional sites across different continents and climate zones would strengthen claims about its broader applicability. Finally, the framework was applied here to two specific devices, and future work should evaluate its performance across a broader range of commercially available and open-source wearable light loggers with different sensor technologies, form factors, and calibration approaches. Despite these limitations, SceneVAL represents a meaningful step toward more ecologically valid and standardised approaches to wearable light logger evaluation and provides a practical and reproducible foundation that other research groups can adopt and extend.

## Conclusion

Wearable light loggers are increasingly used to study the relationship between light exposure and human health, and the validity of the inferences drawn from such studies depends directly on whether these devices accurately capture light exposure under real-world conditions. Yet until now, no standardised framework existed to evaluate their performance in the field. SceneVAL addresses this gap by combining laboratory-grade reference instruments, wearable light loggers, and image-based scene characterisation across naturalistic scenes, providing ground-truth field validation that prior approaches could not achieve. Applied across 433 scenes at two geographically distinct sites, the framework demonstrated that both tested devices performed with high overall agreement but exhibited systematic underestimation that varied meaningfully with lighting condition, time of day, and scene spatial complexity, patterns that would be invisible to laboratory characterisation or wearable-to-wearable comparison alone. Critically, the composition of the validation sample mattered as much as its size: single-condition assessments produced substantially divergent bias estimates, while a diverse selection of approximately 100 scenes was sufficient for stable and representative performance characterisation. Cross-site validation confirmed that the framework is reproducible across independent deployments using locally available instrumentation, supporting its adoption by other research groups without requiring specialised or standardised equipment. Together, these findings establish our framework as a practical and ecologically valid tool for wearable light logger evaluation and provide concrete guidance, on scene diversity, sample size, and analytical approach, for future validation studies seeking to ensure that wearable measurements accurately reflect the light environments they are designed to capture.

## Acknowledgements

### Author contributions

Conceptualization: BG, AD, MS

Methodology: BG, NT, CO, JZ, AD, MS

Data collection: BG, CO

Project administration: AD, MS

Data curation: BG

Software: BG

Formal analysis: BG, AD, MS

Visualization: BG, CO

Validation: BG, AD, MS

Resources: BG, NT, AD, MS

Funding acquisition: BG, AD, MS

Supervision: AD, MS

Writing – original draft: BG

Writing – review and editing: BG, NT, CO, JZ, AD, MS

All authors reviewed and approved the final manuscript.

## Code availability statement

All custom code used for data processing, synchronization, statistical analysis, and figure generation was developed in Python (v3.12) and R (v4.4.3). The complete analysis pipeline and source code are publicly available at: https://github.com/tscnlab/GemiciEtAl_bioRxiv_2026.

## Data availability statement

The datasets generated and analysed during the current study are publicly available together with the analysis code at: https://github.com/tscnlab/GemiciEtAl_bioRxiv_2026.

## Funding

B.G. received support through the Erasmus+ Programme of the European Union. M.S. received funding through the Max Planck Society (Max Planck Research Group), Velux Stiftung (Project No. 1636) and Reality Labs Research. This work was also supported by the MeLiDos project (22NRM05 MeLiDos), which has received funding from the European Partnership on Metrology, co-financed by the European Union's Horizon Europe Research and Innovation Programme and by the Participating States. Views and opinions expressed are however those of the author(s) only and do not necessarily reflect those of the European Union or EURAMET. Neither the European Union nor the granting authority can be held responsible for them.

## Conflict of interest statement

J.Z. declares the following potential conflicts of interest in the past five years (2021–2025): academic roles as Member of Joint Technical Committee 20 (JTC20) of the International Commission on Illumination (CIE), Member of the Research Data Alliance Working Group Optical Radiation and Visual Experience Data, and Speaker of group 2, melanopic effects of light, of the Technical Scientific Committee (TWA) of the German Society of Lighting Technology and Design (LiTG); remunerated roles as examiner for the Swiss Lighting Society; teacher for LiTG, the University of Applied Sciences Munich, and the Technical University of Applied Sciences Rosenheim; associated partner at 3lpi lighting design + engineering, Munich; tool and 3D-model designer for Zumtobel Lighting GmbH; and course designer for the University of Applied Sciences Munich and Virtual University Bavaria; honoraria for talks from LiTG, Lamilux/Heinrich Strunz GmbH, Robert-Bosch Hospital Stuttgart, Ergotopia GmbH, the German statutory accident insurance institution for the administrative sector (VBG), BRIXEN CULTUR, KITEO GmbH & Co. KG, and the University of Applied Sciences Augsburg; travel reimbursements from the Daimler und Benz Stiftung; and, together with 3lpi, holding a design patent for a non-visually optimized luminaire, No. 008194021–0001 through -0006, at the European Union Intellectual Property Office.

A.D. declares the following potential conflicts of interest in the past five years. Academic roles: Member of Joint Technical Committee 20 (JTC20) of the International Commission on Illumination (CIE); Division reporter (DR6-50) of the International Commission on Illumination (CIE) to report outcomes of the “4th Manchester Workshop on Light Metrics for Biology – Light Pollution”. A.D. declares no influence of the disclosed roles or relationships on the work presented herein.

M.S. declares the following potential conflicts of interest in the past five years (2021–2025): academic roles as Member of the Board of Directors of the Society of Light, Rhythms, and Circadian Health (SLRCH), Chair of Joint Technical Committee 20 (JTC20) of the International Commission on Illumination (CIE), Member of the Daylight Academy, and Chair of the Research Data Alliance Working Group Optical Radiation and Visual Experience Data; remunerated roles as Speaker of the Steering Committee of the Daylight Academy, ad-hoc reviewer for the Health and Digital Executive Agency of the European Commission, ad-hoc reviewer for the Swedish Research Council, Associate Editor for LEUKOS, examiner for the University of Manchester, Flinders University, and the University of Southern Norway, and consultant for LyS Technologies and RoX Health; research funding and support from the Max Planck Society, Max Planck Foundation, Max Planck Innovation, Technical University of Munich, Wellcome Trust, National Research Foundation Singapore, European Partnership on Metrology, VELUX Foundation,

Bayerisch-Tschechische Hochschulagentur (BTHA), BayFrance/Bayerisch-Französisches Hochschulzentrum, BayFOR/Bayerische Forschungsallianz, and Reality Labs Research; honoraria for talks from ISGlobal, the Research Foundation of the City University of New York, and the Stadt Ebersberg, Museum Wald und Umwelt; travel reimbursements from the Daimler und Benz Stiftung; and being named on European Patent Application EP23159999.4A, "System and method for corneal-plane physiologically-relevant light logging with an application to personalized light interventions related to health and well-being." M.S. declares that the disclosed roles and relationships had no influence on the work presented herein. The funders had no role in study design, data collection and analysis, the decision to publish, or preparation of the manuscript.

The remaining authors declare no competing interests.